\documentclass[a4paper,twocolumn,showpacs,nofootinbib,floatfix,superscriptaddress,prl]{revtex4}
\usepackage{graphicx}
\usepackage{amsfonts}
\usepackage{amsmath}
\usepackage{amsbsy}
\usepackage{longtable}
\usepackage{bm}
\usepackage{hyperref}
\usepackage{float}

\renewcommand{\vec}[1]{\mathbf{#1}}
\def\be{\begin{equation}}
\def\ee{\end{equation}}
\def\bea{\begin{eqnarray}}
\def\eea{\end{eqnarray}}

\begin{document}

\title{Relativistic shock waves in viscous gluon matter}

\author{I.\ Bouras}
\affiliation{Institut f\"ur Theoretische Physik, 
Johann Wolfgang Goethe-Universit\"at, 
Max-von-Laue-Str.\ 1, D-60438 Frankfurt am Main, Germany}

\author{E.\ Moln\'ar}
\author{H.\ Niemi}
\affiliation{Frankfurt Institute for Advanced Studies, 
Ruth-Moufang-Str.\ 1, D-60438 Frankfurt am Main, Germany}

\author{Z.\ Xu}
\author{A.\ El}
\author{O.\ Fochler}
\author{C.\ Greiner}
\affiliation{Institut f\"ur Theoretische Physik, 
Johann Wolfgang Goethe-Universit\"at, 
Max-von-Laue-Str.\ 1, D-60438 Frankfurt am Main, Germany}

\author{D.H.\ Rischke}
\affiliation{Institut f\"ur Theoretische Physik, 
Johann Wolfgang Goethe-Universit\"at, 
Max-von-Laue-Str.\ 1, D-60438 Frankfurt am Main, Germany}
\affiliation{Frankfurt Institute for Advanced Studies, 
Ruth-Moufang-Str.\ 1, D-60438 Frankfurt am Main, Germany}

\begin{abstract}
We solve the relativistic Riemann problem in viscous 
gluon matter employing a microscopic parton cascade.
We demonstrate the transition from ideal to viscous 
shock waves by varying the shear viscosity to entropy density 
ratio $\eta/s$ from zero to infinity.
We show that an $\eta/s$ ratio larger than 0.2
prevents the development of well-defined shock waves 
on timescales typical for ultrarelativistic heavy-ion collisions.
Comparisons with viscous hydrodynamic calculations 
confirm our findings.

\end{abstract}


\date{\today}

\maketitle


In the 1970's, shock waves were theoretically 
predicted to occur in collisions of heavy nuclei
\cite{Scheid:1974zz}. This phenomenon has been experimentally
investigated \cite{Baumgardt:1975qv} and subsequently
observed \cite{Gutbrod:1989wd}. Recently,  
jet quenching \cite{Adams:2003kv} has been discovered in 
heavy-ion collisions at Brookhaven National Laboratory's
Relativistic Heavy-Ion Collider (RHIC). In this context,
very exciting 
jet-associated particle correlations \cite{Wang:2004kfa} 
have been observed, which indicates the formation of
shock waves in form of Mach cones \cite{Stoecker:2004qu}
induced by supersonic partons moving through the 
quark-gluon plasma (QGP).
If true, it could give a direct access to 
the equation of state of the QGP, because the Mach cone angle
is given by $\alpha = \arccos (c_s/v_{\rm jet})$, where
$c_s$ is the velocity of sound of the QGP.
The velocity of sound is related to
the equation of state via $c_s^2 = dP/de$, 
where $P$ is the pressure and $e$ the energy density.

Shock waves can form and propagate only if
matter behaves like a fluid.
The large measured elliptic flow coefficient 
$v_2$ \cite{Adler:2003kt} indicates that the QGP created
at the RHIC could even be a nearly perfect fluid.
This is confirmed by recent calculations
within viscous hydrodynamics \cite{Luzum:2008cw}
and microscopic transport theory \cite{Xu:2007jv}
which estimate the shear viscosity to entropy density
ratio $\eta/s$ to be less than $0.4$ in order to not spoil
the agreement with the $v_2$ data.
However, it is an important question whether the 
$\eta/s$ value deduced from $v_2$ data is sufficiently small 
to allow for the formation of shock waves.

In this Letter we make an effort to answer this question
by considering the relativistic Riemann problem
in viscous gluon matter. Using the BAMPS microscopic
transport model
(BAMPS denotes the Boltzmann approach of multiparton scatterings) 
\cite{Xu:2004mz} we demonstrate the transition from ideal 
shock waves with zero width, to viscous shock waves with 
nonzero width, to free diffusion by varying the 
shear viscosity to entropy density 
ratio $\eta/s$ from zero to infinity. We estimate the 
upper limit of the $\eta/s$ ratio, for which shocks can 
still be observed experimentally on the time scale of
an ultrarelativistic heavy-ion collision.


The initial condition for the relativistic Riemann problem 
consists of two regions of thermodynamically equilibrated
matter with different constant pressure separated by a membrane
at $z=0$, which is removed at $t=0$. 
\begin{figure}[th]
\includegraphics[width=8cm]{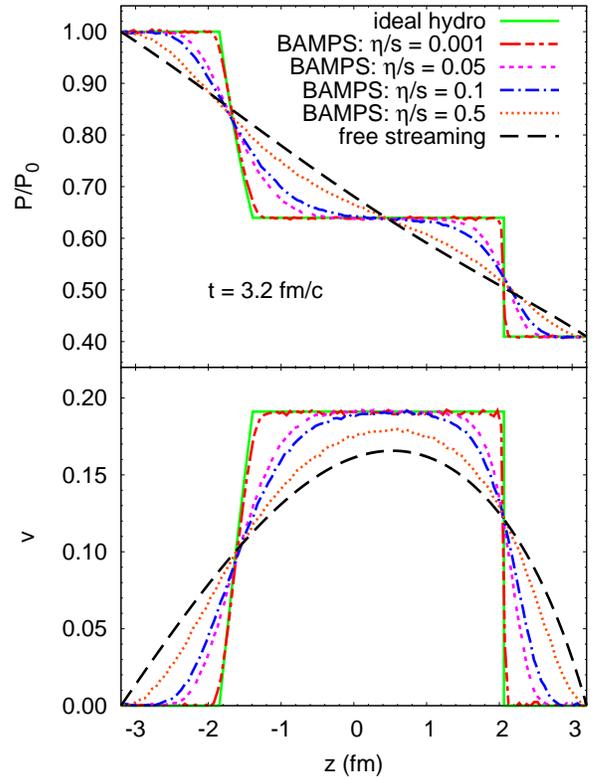}
\caption{(Color online) The solution of the Riemann problem.
At $t=0$, the pressure is $P_0=5.43 \ {\rm GeV fm}^{-3}$ for
$z<0$ and $P_4 =2.22 \ {\rm GeV fm}^{-3}$ for $z>0$.
The upper panel shows the pressure and the lower panel
the velocity at time $t=3.2$ fm/c.
}
\label{fig01}
\end{figure}
Matter is assumed to be homogeneous in the
transverse ($x,y$) direction so that
the further evolution of matter becomes (1 + 1)-dimensional. 
For a perfect fluid, i.e., $\eta=0$,
this problem has an analytical solution
\cite{Schneider:1993gd}, given by the
solid curves in Fig.\ \ref{fig01}. For a thermodynamically normal
medium, 
there is a shock wave travelling into matter with the
lower pressure (on the right) with velocity $v_{\rm shock}$
larger than the velocity of sound $c_s$.
The region of constant pressure behind the shock wave
is the so-called shock plateau. Here, matter 
moves collectively with a
constant velocity $v_{\rm plat}$ as shown in the bottom panel of 
Fig.\ \ref{fig01}. Simultaneously with the creation of 
the shock wave, a rarefaction wave travels with velocity $c_s$
into matter with the larger pressure (on the left).
For an ideal fluid, the solution of the Riemann problem
is self-similar in the variable 
$\zeta=z/t$, i.e., the solution as a function of $\zeta$ 
does not change with time.

The velocity of the shock front is determined from 
the relativistic Rankine-Hugoniot-Taub equations 
\cite{Taub:1948zz} and is given by \cite{Schneider:1993gd}
\begin{equation}
\label{vshock}
v_{\rm shock} = \left [ \frac{(P_3-P_4)(e_3+P_4)}{(e_4+P_3)
(e_3-e_4)} \right ]^{1/2}\, ,
\end{equation}
where $P_3$ and $e_3$ ($P_4$ and $e_4$) 
are the pressure and energy density on the left (right) 
side of the shock front.
For non-interacting ultrarelativistic gluon matter 
the equation of state is $e=3P$. For the situation depicted
in Fig.\ \ref{fig01}, $v_{\rm shock}= 0.645$.

The extreme opposite of an ideal fluid (i.e., $\eta = 0$)
is a gas of free-streaming particles (i.e., $\eta=\infty$). 
The solution of the Riemann problem for free-streaming particles
is given by the short-dashed lines in Fig.\ \ref{fig01},
which can also be calculated analytically \cite{Greiner:1996md}.
From the solution for the single-particle distribution function 
$f(x,p)$, one computes the energy-momentum tensor 
$T^{\mu \nu}= \int (d^3p/E)\, p^\mu p^\nu \,f(x,p)$, where 
$p^{\mu} = (E,\vec{p})$ is the four-momentum.
The energy density defined as $e=u_\mu T^{\mu \nu} u_\nu$ is the 
largest eigenvalue of $T^\mu_\nu$. The eigenvector $u^\mu$ is then
the four velocity in Landau's definition [14]. In our case
$u^\mu=\gamma(1,0,0,v)$, where $v=T^{03}/(e+T^{33})$ and
$\gamma = (1-v^2)^{-1/2}$.
The pressure is defined as $P = -\Delta_{\mu \nu} T^{\mu \nu}/3$,
where $\Delta^{\mu \nu}=g^{\mu \nu}-u^\mu u^\nu$ and 
$g^{\mu \nu}=(1,-1,-1,-1)$ is the metric tensor.
For systems of massless particles $T^\mu_\nu$ is traceless and
thus $P=e/3$.
In the free-streaming case the characteristic structure of
the solution of the Riemann problem for an ideal fluid
is completely washed out, and a clear distinction between 
the shock wave and the rarefaction fan is no longer possible.


In the following, we study the influence of the $\eta/s$ ratio 
on the formation and evolution of shock waves by
solving the Riemann problem with the parton cascade BAMPS 
\cite{Xu:2004mz}. BAMPS is a microscopic transport model 
which solves the Boltzmann equation
$p^{\mu} \partial_{\mu} f(x,p) = C(x,p)$
for on-shell gluons with the collision integral $C(x,p)$.
In this study, we consider only binary gluon scattering
processes with an isotropic cross section.
We remark that, although perturbative QCD (pQCD) favors
small angles in binary gluon scatterings,
the use of an isotropic cross section effectively
implements pQCD gluon bremsstrahlung,
which has a broader angular distribution due to the suppression
of soft collinear radiation \cite{Xu:2004mz}. 

In our calculations we use a constant $\eta/s$ value. 
In order to achieve this, we have to locally adjust the 
cross section $\sigma$, since the particle density $n$ is 
not constant. The shear viscosity 
$\eta$ is given by $\eta = 4 e/(15R^{\mathrm{tr}})$ 
\cite{Xu:2007ns}, where the transport 
collision rate $R^{\mathrm{tr}} = n \sigma^{\mathrm{tr}}=
2n \sigma/3$ for isotropic scattering processes 
\cite{Huovinen:2008te}.
$n$ is calculated via $n = N^\mu u_\mu$, where
$N^\mu = \int (d^3p/E) p^\mu \, f(x,p)$.
We obtain 
\begin{equation}
\label{eta}
\eta=\frac{2}{5} e \lambda_{\rm mfp} \,,
\end{equation}
where $\lambda_{\rm mfp}=1/(n\sigma)$ is the gluon 
mean-free path. Binary collisions imply that we cannot
maintain chemical equilibrium. In this case,
in kinetic equilibrium the entropy density is 
calculated approximately via 
$s = 4n - n \ln \lambda$, where 
$\lambda=n/n_{\mathrm{eq}}$ is the gluon fugacity measuring
the deviation from chemical equilibrium.
For a nonvanishing shear viscosity, the system will 
deviate from initial kinetic and chemical 
equilibrium during the evolution \cite{El:2008yy}.
Gluons are regarded as Boltzmann particles, so that 
the number density in thermal equilibrium is given by 
$n_{\mathrm{eq}}=d_G T^3/\pi^2$
with $d_G=16$ for gluons and $T=e/(3n)$ denotes the temperature. 
Sending $\eta/s$ to zero the cross section will be 
unphysically large. However, it serves as a useful test of the 
parton cascade in the ideal hydrodynamical limit.

In Fig.\ \ref{fig01}, we show the results for various 
$\eta/s$ values as computed with BAMPS,
demonstrating the gradual transition from the 
ideal hydrodynamical limit to free streaming of particles.
Remarkably, the ideal solution is reproduced to
very high precision by the BAMPS calculation for a small
$\eta/s=0.001$. In this sense, BAMPS can compare
with state-of-the-art numerical algorithms used to solve the 
ideal hydrodynamical equations 
\cite{Schneider:1993gd}.
A larger $\eta/s$ value results in a finite transition 
layer where the quantities change smoothly rather than 
discontinuously as in the case of a perfect fluid. 
The width of the shock front is proportional 
to the shear viscosity
\cite{Csernai_book}.

\begin{figure}[ht]
\includegraphics[width=7cm]{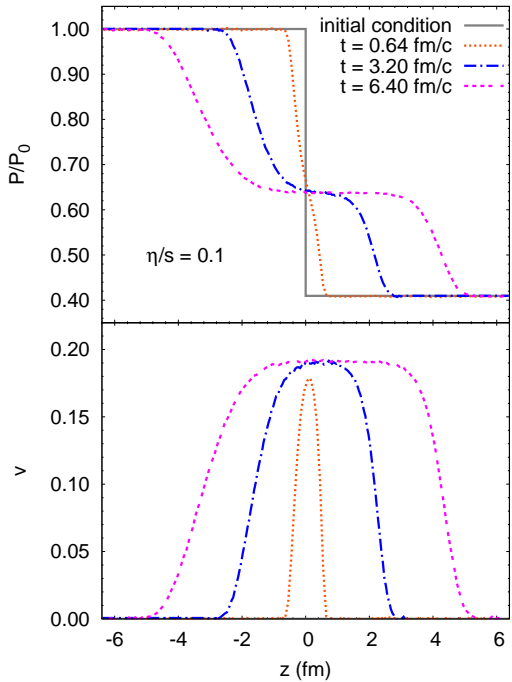}
\caption{(Color online) The time evolution of the
Riemann problem for $\eta/s=0.1$ and the initial condition
of Fig.\ \ref{fig01}.}
\label{fig02}
\end{figure}

As seen in Fig.\ \ref{fig01}, a nonzero viscosity
smears the pressure and velocity profiles and
impedes a clean separation of
the shock front from the rarefaction fan. 
A criterion for a clear separation, and thus the
observability of a shock wave,
is the formation of a shock plateau,
as in the ideal-fluid case. The formation of a shock plateau
takes a certain amount of time, as demonstrated in 
Fig.\ \ref{fig02} where
we show the pressure and velocity profiles at 
different times for $\eta/s=0.1$.
Formally, we define the time of formation of the
shock plateau when the
maximum of the velocity distribution $v(z)$ reaches the value 
$v_{\rm plat}$ of the ideal-fluid solution.
From the bottom part of Fig.\ \ref{fig02}, we
see that this happens at $t=3.2$ fm/c. 
This agrees with what we obtained in the bottom part of 
Fig.\ \ref{fig01}. From this figure we also infer that, for
$\eta/s>0.1$, a shock plateau has not yet developed at
$t=3.2$ fm/c, whereas for $\eta/s <0.1$, it has already fully
formed. 


In order to understand the timescale of formation of
a shock wave, we define the quantity 
\cite{Bhalerao:2005mm,Huovinen:2008te,El:2008yy}
\begin{equation}
\label{knudsen}
K = \frac{\lambda_{\rm mfp} }{L}\,,
\end{equation}
where $L \equiv t \, (v_{\rm shock}+c_s)$ is
the width of the region bounded by the rarefaction
wave travelling to the left and the shock front moving
to the right (in the ideal-fluid case).
Note that $\lambda_{\rm mfp}$ is not constant;
we approximate it by its maximum value which is assumed
on the low-pressure side (the undisturbed medium)
in front of the shock wave. The quantity $K$ can be viewed
as a ``global'' Knudsen number for the Riemann problem.
$K$ goes to zero at late times, which implies
that the medium behaves
more and more like an ideal system.\\
We find a scaling behavior for the solution of the Riemann 
problem: 
The pressure profile $P(z,t;\eta/s)/P_0$ (and also the 
velocity profile) is only a function of $\zeta=z/t$ and $K$,
i.e., $P(z,t;\eta/s)/P_0=F(\zeta; K)$, for a given ratio 
$P_4/P_0$.
The value of $K$ at which the shock wave forms, called $K_f$,
is universal for a given $P_4/P_0$. With Eq.\ (\ref{eta}) 
we obtain $\lambda_{\rm mfp}=10/(3T_4) \eta/s$ in the 
undisturbed medium with the lower pressure.
Thus, using the observation from Fig.\ \ref{fig02} that
the shock wave forms at $t=3.2$ fm/c for $\eta/s=0.1$,
we find that $K_f=0.053$.

\begin{figure}[t]
\includegraphics[width=7.0cm]{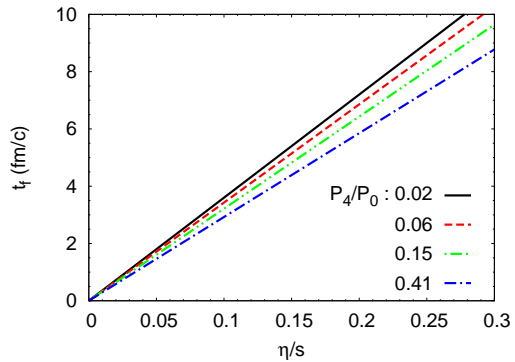}
\caption{(Color online) Formation time of shock waves 
as a function of $\eta/s$.}
\label{fig03}
\end{figure}

By inserting $\lambda_{\rm mfp}=10/(3T) \eta/s$ into 
Eq.\ (\ref{knudsen}),
the formation time of shock waves is given by
\begin{equation}
\label{timeofform}
t_f=\frac{10}{3} \, \frac{1}{K_f (v_{\rm shock}+c_s) T} \, 
\frac{\eta}{s}\,.
\end{equation}
Figure \ref{fig03} shows the relation (\ref{timeofform}) 
with $T=350$ MeV and for
various initial pressure discontinuities $P_4/P_0$.
The difference in slopes reflects the dependence of $K_f$ and 
$v_{\rm shock}$ on the ratio $P_4/P_0$.
For $\eta/s=0.2$ a shock will not be visible until 
6$-$7.2 fm/c, which most likely
exceeds the lifetime of the QGP at the RHIC. 
Vice versa, if shock phenomena are discovered at RHIC, 
this could be an indication that the QGP has
a small $\eta/s$ ratio, probably smaller than $0.2$. 
For a more viscous QGP no shock waves
and thus no Mach cones will be formed. 
In a relativistic heavy-ion
collision, the temperature is decreasing during the expansion. 
Thus, according to Eq.\ (\ref{timeofform}), $t_f$ is even
larger and shock waves may be even harder to observe.


To confirm our results calculated with BAMPS, we solve the
Riemann problem within relativistic dissipative fluid dynamics.
To this end, we use the Israel-Stewart (IS) equations 
\cite{Israel_Stewart} (see
Ref.\ \cite{Denicol:2008rj} for different approaches).
We neglect heat conductivity and bulk viscosity (thus 
assuming local chemical equilibrium).
In 1+1 dimensions, the IS equations reduce to
\begin{eqnarray}
\label{cons_e}
&&\partial_t T^{00} + \partial_z (v T^{00})\!\! = 
\!\!- \partial_z (v P + v \tilde \pi) , \\
\label{cons_m}
&& \partial_{t} T^{0z} + \partial_z (v T^{0z}) \!\! =
\! \! - \partial_z (P + \tilde \pi) , \\
\label{cons_pi}
&&\gamma \partial_{t} \tilde \pi + \gamma v \partial_z \tilde 
\pi \! \!= \! \!\frac{1}{\tau_{\pi}} \!\left(\pi_{NS} - 
\tilde \pi \right) - \frac{\tilde \pi}{2} \! 
\left(\!\theta + D \ln \frac{\beta_2}{T}\!\right) ,
\end{eqnarray}
where $\theta \equiv \partial_{\mu} u^{\mu}$ and 
$D \equiv u^{\mu} \partial_{\mu}$.
The laboratory frame energy and momentum 
density are given by $T^{00} = (e + P + \tilde \pi)\gamma^2 
- (P + \tilde \pi)$,
and $T^{0z} = (e + P +\tilde{\pi})\gamma^2 v$, 
while the system is closed by a massless gluon equation of state.
For vanishing shear viscous pressure, $\tilde \pi$, Eqs.\ 
(\ref{cons_e}) and (\ref{cons_m}) reduce to the equations 
of ideal hydrodynamics.\\
The first-order theory of relativistic dissipative 
fluid dynamics defines the viscous pressure algebraically 
by the relativistic Navier-Stokes value
$\pi_{NS} = -(4/3)\eta \theta$ \cite{Eckart:1940te,Landau_book}.
In the IS equations, the local shear pressure relaxes to the 
Navier-Stokes value on a timescale comparable to the 
mean-free time between collisions, given 
by the relaxation time $\tau_{\pi} = 2\eta \beta_2$, 
where $\beta_2 = 3/(4P)$ \cite{Israel_Stewart}.

\begin{figure}[t]
\includegraphics[width=7.0cm]{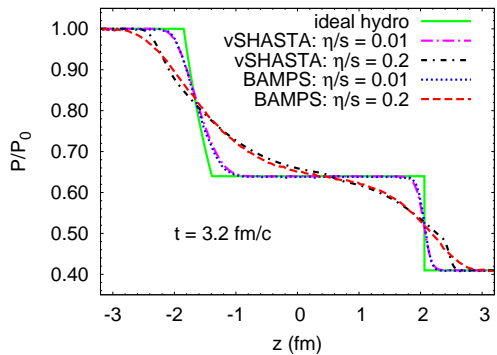}
\caption{(Color online) Same as Fig.\ \ref{fig01}. Results
obtained from vSHASTA and BAMPS calculations are compared.
}
\label{fig04}
\end{figure}

The IS equations (\ref{cons_e})-(\ref{cons_pi}) are solved 
numerically using the
viscous sharp and smooth transport algorithm
(vSHASTA) approach \cite{Molnar:2008fv}.
Figure \ref{fig04} shows comparisons
between the results from BAMPS and vSHASTA calculations for
$\eta/s = 0.01$ and $0.2$.
We see perfect agreement for $\eta/s = 0.01$, whereas for 
the larger value $\eta/s = 0.2$ small deviations in the 
region around the shock front and 
the rarefaction fan are found.  
The reason can be understood considering
the Knudsen number $K_\theta \equiv 
\lambda_{\rm mfp} \theta$. The IS equations contain
terms of second order in $K_\theta$ \cite{Betz:2008me}.
However, at the shock front, the macroscopic scale over which
the fluid velocity varies is comparable to the
microscopic scale $\lambda_{\rm mfp}$, such that 
$K_\theta \sim 1$,  and one would need higher
powers of $K_\theta$ in the IS equations to improve the
description. Wherever $K_\theta$ is large, the applicability of
the IS theory is questionable. Microscopic transport calculations
do not suffer from this drawback.

In summary, employing the parton cascade BAMPS 
we have solved the relativistic Riemann problem. 
The transition from ideal-fluid behavior to free streaming
is demonstrated. 
Numerical results from BAMPS agree well
with those obtained from 
viscous hydrodynamical calculations.
We found that the formation of shock waves
in gluon matter with $\eta/s > 0.2$ probably takes longer
than the lifetime of the QGP at the RHIC.
Whether Mach cones from jets \cite{Neufeld:2008dx}
can be observed within nuclear collisions 
will be studied in the future within the BAMPS approach.

The authors are grateful to B.\ Betz, L.P.\ Csernai, J.A.\ 
Maruhn, H.\ St\"ocker, and G.\ Torrieri for discussions 
and to the Center for the Scientific 
Computing (CSC) at Frankfurt for the computing resources.
E.\ M. acknowledges support by the Alexander von
Humboldt foundation. The work of H.\ N. was supported by
the Extreme Matter Institute (EMMI).
This work was supported by the Helmholtz International Center
for FAIR within the framework of the LOEWE program 
launched by the State of Hesse.




\begin{thebibliography}{99}
\bibitem{Scheid:1974zz}
  W.~Scheid, H.~Muller and W.~Greiner,
  Phys.\ Rev.\ Lett.\  {\bf 32} (1974) 741.
\bibitem{Baumgardt:1975qv}
  H.~G.~Baumgardt {\it et al.},
  Z.\ Phys.\  A {\bf 273} (1975) 359.
\bibitem{Gutbrod:1989wd}
  H.~H.~Gutbrod, A.~M.~Poskanzer and H.~G.~Ritter,
  Rept.\ Prog.\ Phys.\  {\bf 52}, 1267 (1989);
  H.~H.~Gutbrod {\it et al.},
  Phys.\ Rev.\  C {\bf 42}, 640 (1990).
\bibitem{Adams:2003kv}
  J.~Adams {\it et al.}  [STAR Collaboration],
  Phys.\ Rev.\ Lett.\  {\bf 91}, 172302 (2003);
  A.~Adare {\it et al.}  [PHENIX Collaboration],
  {\it ibid.} {\bf 101}, 232301 (2008).
\bibitem{Wang:2004kfa}
  F.~Wang  [STAR Collaboration],
  J.\ Phys.\ G {\bf 30}, S1299 (2004);
  J.~Adams {\it et al.}  [STAR Collaboration],
  Phys.\ Rev.\ Lett.\  {\bf 95}, 152301 (2005);
  S.~S.~Adler {\it et al.}  [PHENIX Collaboration],
  {\it ibid.} {\bf 97}, 052301 (2006);
  J.~G.~Ulery  [STAR Collaboration],
  Nucl.\ Phys.\  A {\bf 774}, 581 (2006);
  N.~N.~Ajitanand  [PHENIX Collaboration],
  {\it ibid.} {\bf 783}, 519 (2007);
  A.~Adare {\it et al.}  [PHENIX Collaboration],
  Phys.\ Rev.\  C {\bf 78}, 014901 (2008).
\bibitem{Stoecker:2004qu}
  H.~St\"ocker,
  Nucl.\ Phys.\  A {\bf 750}, 121 (2005);
  J.~Ruppert and B.~M\"uller,
  Phys.\ Lett.\  B {\bf 618}, 123 (2005);
  J.~Casalderrey-Solana, E.~V.~Shuryak and D.~Teaney,
  J.\ Phys.\ Conf.\ Ser.\  {\bf 27}, 22 (2005);
  V.~Koch, A.~Majumder and X.~N.~Wang,
  Phys.\ Rev.\ Lett.\  {\bf 96}, 172302 (2006).
\bibitem{Adler:2003kt}
  S.~S.~Adler {\it et al.}  [PHENIX Collaboration],
  Phys.\ Rev.\ Lett.\  {\bf 91}, 182301 (2003);
  J.~Adams {\it et al.}  [STAR Collaboration],
  {\it ibid.} {\bf 92}, 052302 (2004);
  B.~B.~Back {\it et al.}  [PHOBOS Collaboration],
  Phys.\ Rev.\  C {\bf 72}, 051901 (2005).
\bibitem{Luzum:2008cw}
  M.~Luzum and P.~Romatschke,
  Phys.\ Rev.\  C {\bf 78}, 034915 (2008);
  H.~Song and U.~W.~Heinz,
  arXiv:0812.4274.
\bibitem{Xu:2007jv}
  Z.~Xu, C.~Greiner and H.~St\"ocker,
  Phys.\ Rev.\ Lett.\  {\bf 101}, 082302 (2008);
  Z.~Xu and C.~Greiner,
  Phys.\ Rev.\  C {\bf 79} (2009) 014904
\bibitem{Xu:2004mz}
  Z.~Xu and C.~Greiner,
  Phys.\ Rev.\  C {\bf 71} (2005) 064901;
  {\bf 76} (2007) 024911.
\bibitem{Schneider:1993gd}
  V.~Schneider {\it et al.},
  J.\ Comput.\ Phys.\  {\bf 105} (1993) 92;
  D.~H.~Rischke, S.~Bernard and J.~A.~Maruhn,
  Nucl.\ Phys.\  A {\bf 595} (1995) 346;
  D.~H.~Rischke,
  arXiv:nucl-th/9809044.
\bibitem{Taub:1948zz}
  A.~H.~Taub,
  Phys.\ Rev.\  {\bf 74} (1948) 328.
\bibitem{Greiner:1996md}
  C.~Greiner and D.~H.~Rischke,
  Phys.\ Rev.\  C {\bf 54}, 1360 (1996).
\bibitem{Landau_book}
L. D. Landau and E. M. Lifshitz,
{\it Fluid Dynamics}, Second Edition, Butterworth-Heinemann (1987).
\bibitem{Xu:2007ns}
  Z.~Xu and C.~Greiner,
  Phys.\ Rev.\ Lett.\  {\bf 100} (2008) 172301.
\bibitem{Huovinen:2008te}
  P.~Huovinen and D.~Molnar,
  Phys.\ Rev.\  C {\bf 79} (2009) 014906
\bibitem{El:2008yy} 
  A.~El {\it et al.},
  Phys.\ Rev.\ C {\bf 79} (2009) 044914
\bibitem{Csernai_book}
L. P. Csernai, {\it Introduction to Relativistic Heavy Ion Collisions}, Wiley (1994);
C. Cercignani and G. M. Kremer,
{\it The Relativisitic Boltzmann Equation: Theory and Applications}, Birkh\"auser (2002);
  J.~A.~S.~Lima and A.~Kandus,
  Phys.\ Rev.\  D {\bf 67} (2003) 023002.
\bibitem{Bhalerao:2005mm}
  R.~S.~Bhalerao {\it et al.},
  Phys.\ Lett.\  B {\bf 627}, 49 (2005).
\bibitem{Israel_Stewart}
W.~Israel,
Annals Phys.\  {\bf 100}, 310 (1976);
J.~M.~Stewart, Proc.\ Roy.\ Soc.\ A {\bf 357}, 59 (1977);
W.~Israel and J.~M.~Stewart,
Annals Phys.\  {\bf 118}, 341 (1979).
\bibitem{Denicol:2008rj}
  G.~S.~Denicol {\it et al.},
  Phys.\ Rev.\  C {\bf 78}, 034901 (2008);
  T.~S.~Biro, E.~Molnar and P.~Van,
  {\it ibid.}  {\bf 78}, 014909 (2008).
\bibitem{Eckart:1940te}
  C.~Eckart,
  Phys.\ Rev.\  {\bf 58} (1940) 919.
\bibitem{Molnar:2008fv}
  E.~Molnar,
  Eur.\ Phys.\ J.\  C {\bf 60} (2009) 413
\bibitem{Betz:2008me}
  B.~Betz, D.~Henkel and D.~H.~Rischke,
  arXiv:0812.1440.
\bibitem{Neufeld:2008dx}
  R.~B.~Neufeld,
  Phys.\ Rev.\  C {\bf 79} (2009) 054909
  B.~Betz {\it et al.},	
  arXiv:0812.4401.

\end{thebibliography}
\end{document}